\documentclass[aps,prb,twocolumn,superscriptaddress,raggedbottom,showpacs,floatfix]{revtex4-1}
\usepackage{amsmath,amsfonts,amssymb,amsthm}
\usepackage{color}
\usepackage{graphicx,latexsym}

\begin{document}
\renewcommand{\vec}{\mathbf}
\renewcommand{\Re}{\mathop{\mathrm{Re}}\nolimits}
\renewcommand{\Im}{\mathop{\mathrm{Im}}\nolimits}

\title{Drag effect and Cooper electron-hole pair fluctuations in a topological insulator film}
\author{D.K. Efimkin}
\affiliation{Institute of Spectroscopy  RAS, 142190, Troitsk, Moscow, Russia}
\author{Yu.E. Lozovik}
\altaffiliation{email: lozovik@isan.troitsk.ru}

\affiliation{Institute of Spectroscopy RAS, 142190, Troitsk, Moscow, Russia}
\affiliation{Moscow Institute of Physics and Technology, 141700,
Moscow, Russia}
\affiliation{Moscow Institute for Electronics and Mathematics at National Research University HSE, 109028, Moscow, Russia}

\begin{abstract}
Manifestations of fluctuating Cooper pairs formed by electrons and holes populating opposite surfaces of a topological insulator film in the Coulomb drag effect are considered. Fluctuational Aslamazov-Larkin contribution to the transresistance between surfaces of the film is calculated. The contribution is the most singular one in the vicinity of critical temperature $T_\mathrm{d}$ and diverges in the critical manner as $(T-T_\mathrm{d})^{-1}$. In the realistic conditions $\gamma\sim T_\mathrm{d}$, where $\gamma$ is average scattering rate of electrons and holes, Aslamazov-Larkin contribution plays important role and can dominate the fluctuation transport. The macroscopic theory based on time-dependent Ginzburg-Landau equation is developed for description of the fluctuational drag effect in the system. The results can be easily generalized for other realizations of electron-hole bilayer.
\end{abstract}
\maketitle

\section{Introduction}
In double layer structures many-body physics can be probed in Coulomb drag effect (see \cite{Rojo} and references therein). Due to a momentum exchange between the charge carriers from different layers an electric current $I_{\mathrm{drive}}$ induced in an active layer leads to electric current in a passive one. If the passive layer is closed $I_{\mathrm{drive}}$ leads to a voltage drop $V_{\mathrm{drag}}$ in it compensating the drag force. In experiments transresistance of the bilayer $\rho_{\mathrm{d}}=V_{\mathrm{drag}}/I_{\mathrm{drive}}$ is measured. If charge carriers from different layers can be considered as weakly coupled Fermi liquids the temperature dependence of a transresistance in wide range of temperatures is $\rho_{\mathrm{D}}\sim T^2$ that has been established both experimentally \cite{DragExp1,DragExp2} and theoretically \cite{ZhengMacDonald,JauhoSmith,KamenevOreg}. Deviation of a transresistance dependence from this usual one can reveal existence of new broken symmetry phases or strong interlayer correlations in a bilayer.

In a system of spatially separated electrons and holes the Coulomb attraction between them can lead to electron-hole Cooper pairing\cite{LozovikYudson}. It has been predicted in a semiconductor heterostructure\cite{LozovikYudson} in graphene double layer system\cite{LozovikSokolik,MinBistrizerSuMacDonald,KharitonovEfetov,AbergelSensarmaDasSarma,PeraliNeilsonHamilton,SuprunenkoCheianovFalko} and in thin film of topological insulator\cite{EfimkinLozovikSokolik,SeradjehMooreFranz,TilahunLeeHAnkiewiczMacDonald}. Electron-hole Cooper pairing can lead to superfluidity \cite{LozovikYudson,BalatskyJoglekarLittlewood}, nonlocal Andreev reflection \cite{PesinMacDonald}, internal Josephson effect \cite{LozovikPoushnov,SternGirvinMacDonaldMa,FoglerWilczek,BezuglyjShevchenko,JosehsonExp} and to strong Coulomb drag effect\cite{VignaleMacDonald}. Particulary rapid increasing of transresistance below critical temperature and its jump at the temperature of Berezinskii-Kosterlitz-Thouless \cite{Berezinskii,KosterlizThouless} transition to the superfluid state have been predicted.

In an electron-hole bilayer Cooper pairs can appear above critical temperature as thermodynamic fluctuations. They can lead to critical behavior of tunneling conductivity that can be interpreted as fluctuational internal Josephson effect\cite{EfimkinLozovik} and to a pseudogap formation in single-particle density of states of electrons and holes\cite{Rist}. Considerable enhancement of drag resistivity by Cooper pair fluctuations in vicinity of the critical temperature that smooths the jump has been predicted\cite{Hu}. In that work Maki-Thompson \cite{Maki,Thompson} (MT) contribution to transresistance, that logarithmically diverges in vicinity of the critical temperature, has been calculated. Later the same dependence has been obtained within kinetic equation approach. In that approach the Coulomb interaction between electrons and holes was renormalized by Cooper pair fluctuations and treated perturbatively\cite{Mink,Vignale}. Hence that contribution to the transresistivity has the single-particle origin.  There is another, Aslamazov-Larkin\cite{AslamazovLarkin} (AL) contribution to the transresistivity that was neglected in those works. Here we calculate it both microscopically and within the macroscopic approach based on time-dependent Ginzburg-Landau equation. The contribution has collective origin and comes from the possibility of fluctuating Cooper pairs to carry electric currents both in electron and hole layers. The fluctuational drag effect is considered for topological insulator thin film but the results can be easily generalized for other realizations of an electron-hole bilayer.

The system of spatially separated composite electrons and composite holes is realized in quantum Hall bilayer at total filling factor $\nu_\mathrm{T}=1$\cite{EisensteinMacDonald}. For that system fluctuational contributions to transresistivity including AL one have been calculated\cite{UssishkinStern,ZhouKim}. But in quantum Hall bilayer interaction between electron and hole is not only Coulomb in origin but also comes from fluctuations of Chern-Simons field. The later is important and influences both drag in weak coupling regime\cite{UssishkinStern2,Sakhi} and fluctuational drag in vicinity of the critical temperature.

The rest of the paper is organized as follows. In the Section 2 we briefly discuss the model. The section 3 is devoted to microscopical description of the Cooper pair fluctuations. In section 4 we present microscopical approach for fluctuational drag effect. In Section 5 we present macroscopical theory of the fluctuational transport. The Section 6 is devoted to analysis of results and discussions.
\section{The model}
Let us consider the system of spatially separated Dirac electrons and holes populating opposite surfaces of topological insulator film. Possibility and the peculiarities of electron-hole Cooper pairing in that system is discussed in details in our paper \cite{EfimkinLozovikSokolik}.  The Hamiltonian of the system in the single-band approximation that ignores valence (conduction) band on the surface with excess of electrons (holes) is given by

\begin{equation}
\begin{split}
H_{\mathrm{eh}}=&\sum_{\vec{p}}\xi_{\vec{p}} a_\vec{p}^+a_\vec{p} - \sum_{\vec{p}}\xi_{\vec{p}} b_\vec{p}^+b_\vec{p} + \\ &+\sum_{\vec{p}\vec{p}'\vec{q}} U(\vec{q})\Lambda_{\vec{p}'-\vec{q},\vec{p}'}^{\vec{p}+\vec{q},\vec{p}}a_{\vec{p}+\vec{q}}^+b^+_{\vec{p}\prime-\vec{q}}
b_{\vec{p}\prime}^{\mathstrut} a_{\vec{p}}^{\mathstrut}.
\end{split}
\end{equation}
Here $a_{\vec{p}}$ is annihilation operator for a electron on the surface with excess of electrons and  $b_{\vec{p}}$ is annihilation operator for a electron on the surface with excess of holes \cite{Comment2}; $\xi_{\vec{p}}=v_{\mathrm{F}}p-E_{\mathrm{F}}$ is Dirac dispersion law in which $v_{\mathrm{F}}$ and $E_{\mathrm{F}}$ are velocity and Fermi energy of electrons and holes. The balanced case is considered since Cooper pairing is sensitive to concentration mismatch of electrons and holes. $U(\vec{q})$ is screened Coulomb interaction between electrons and holes (see \cite{EfimkinLozovikSokolik} for its explicit value) and $\Lambda_{\vec{p}'-\vec{q},\vec{p}'}^{\vec{p}+\vec{q},\vec{p}} =
\cos{ (\phi_{\vec{p},\vec{p}+\vec{q}}/2)} \cos{(\phi_{\vec{p}',\vec{p}'+\vec{q}}/2)}$ is angle factor originating from the overlap of spinor wave functions of two-dimensional Dirac fermions. Critical temperature of pairing in weak-coupling or Bardeen-Cooper-Schrieffer (BCS) regime is given by
\begin{equation}
T_0=\frac{2\gamma'E_{\mathrm{F}}}{\pi} e^{-1/\nu_\mathrm{F}U'},
\end{equation}
where $\gamma'=e^C$ where $C\approx0,577$ is the Euler constant; $\nu_\mathrm{F}=E_\mathrm{F}/2\pi v_\mathrm{F}^2$ is density of states of electrons and holes on Fermi level. Here $U'$ is Coulomb coupling constant. In our work\cite{EfimkinLozovikSokolik} we have calculated Coulomb coupling constant in static limit of Random Phase Approximation. The maximal value of dimensionless coupling constant for realistic $\hbox{Bi}_2\hbox{Se}_3$ TI films can achieve $\nu_\mathrm{F}U'\approx0.18$ that corresponds to $T_0\sim0.1\;\hbox{K}$. But we aware that this approximation can considerably underestimate critical temperature. But the dynamical and multiband effects \cite{LozovikOgarkovSokolik,LozovikSokolikMultiband,SodemannPesinMacDonald} can be incorporated in our theory by renormalization of Coulomb coupling constant, so here we treat $T_0$ as phenomenological parameter.

We do not specify explicitly the interaction Hamiltonian with disorder. Since components of a Cooper pair are spatially separated and have opposite charge both short-range disorder and long-range Coulomb impurities lead to the pairbreaking and can suppress Cooper pairing. Below we introduce phenomenological scattering rates of electrons $\gamma_\mathrm{e}$ and holes $\gamma_\mathrm{h}$.

\section{The Cooper pair fluctuations}
For microscopical description of gaussian Cooper pair fluctuations we introduce Cooper propagator $\Gamma_\mathrm{c}^{\mathrm{R}}(\omega,\vec{q})$\cite{LarkinVarlamov}. It corresponds to the two-particle vertex function in the Cooper channel and satisfies the Bethe-Salpeter equation depicted on Fig.1-a. In BCS regime its solution can be presented in the form
\begin{equation}\label{CooperPropagator1}
\Gamma_\mathrm{c}^{\mathrm{R}}(\omega,\vec{q})=\frac{U'}{1-U'\Pi_{\mathrm{c}}^{\mathrm{R}}(\omega,\vec{q})},
\end{equation}
where $\Pi_{\mathrm{c}}^{\mathrm{R}}(\omega,\vec{q})$ corresponds to electron-hole bubble diagram. It can be interpreted as Cooper susceptibility of the system and can be analytically continued from
\begin{equation}
\label{CooperSusceptibility}
\Pi_{\mathrm{c}}(i\Omega_n,\vec{q})=-T\sum_{\vec{p}\vec{\omega_n}}G_\vec{e}(i\omega_n+i\Omega_n,\vec{p}+\vec{q})G_\mathrm{h}(i\omega_n,\vec{p}),
\end{equation}
where $G_{\mathrm{e}(\mathrm{h})}(i\omega_n,\vec{p})=(i\omega_m\mp\xi_{p}+i\gamma_{\mathrm{e}(\mathrm{h})} \mathrm{sgn} \omega_n) $ is the single-particle Green function of the electrons in the corresponding layer and $\gamma_{\mathrm{e}(\mathrm{h})}$ is phenomenologically introduced scattering rate. Frequencies $\Omega_n=2\pi n T$ and $\omega_n=(2n+1) \pi T$ are bosonic and fermionic Matsubara ones. After direct evaluation of (\ref{CooperSusceptibility}) we obtain for the Cooper propagator
\begin{equation}
\label{CooperPropagator2}
\begin{split}
\Gamma_\mathrm{c}^{\mathrm{R}}(\omega,\vec{q})=(\nu_{\mathrm{F}}\left(\ln\frac{T}{T_\mathrm{0}}+\Psi\left(\frac{1}{2}+\frac{\gamma}{2\pi T}\right)-\Psi\left(\frac{1}{2}\right)\right) - \\ -  i\frac{\nu_\mathrm{F}\omega}{4 \pi T}\Psi^\prime\left(\frac{1}{2}+\frac{\gamma}{2\pi T}\right) + \frac{\nu_{\mathrm{F}} v_{\mathrm{F}}^2\vec{q}^2}{64\pi^2T^2} \left|\Psi^{\prime\prime}\left(\frac{1}{2}+\frac{\gamma}{2\pi T}\right)\right|)^{-1}.
\end{split}
\end{equation}
Here  $\Psi(x)$ is the digamma function and $\gamma=(\gamma_\mathrm{e}+\gamma_{\mathrm{h}})/2$ is disorder caused Copper pair scattering rate\cite{Comment4}. In the absence of disorder $\Gamma_\mathrm{c}^{\mathrm{R}}(0,0)=0$ at the critical temperature $T_0$ indicating Cooper instability of the system toward Cooper pairing. Critical temperature for disordered system $T_\mathrm{d}$ at which $\Gamma_\mathrm{c}^{\mathrm{R}}(0,0)=0$ satisfies the following equation
\begin{equation}
\ln\frac{T_\mathrm{d}}{T_\mathrm{0}}+\Psi\left(\frac{1}{2}+\frac{\gamma}{2\pi T}\right)-\Psi \left(\frac{1}{2}\right)=0.
\end{equation}
This equation has nontrivial solution if $\gamma<0.89 T_0$. In the opposite case the pairing is suppressed by disorder. Below we suppose that the Cooper pairing is not suppressed by disorder.

Above the critical temperature $T_{\mathrm{d}}$ the expression for Cooper pair propagator (\ref{CooperPropagator2}) can be approximated in the following way
\begin{equation}
\Gamma_\mathrm{c}^{\mathrm{R}}(\omega,\vec{q})=-\frac{1}{i\gamma_\mathrm{c} \omega -\epsilon_{\vec{q}}}.
\end{equation}
where $\epsilon_{\vec{q}}=a+D\vec{q}^2$ can be interpreted as energy of fluctuating Cooper pair and the corresponding coefficients are given by
\begin{equation}
\begin{split}
\label{Coefficients}
a=\nu_\mathrm{F}\ln\frac{T}{T_{\mathrm{d}}}, \quad \quad \gamma_{\mathrm{c}}=\frac{\nu_\mathrm{F}}{4 \pi T_{\mathrm{d}}}\Psi^\prime\left(\frac{1}{2}+\frac{\gamma}{2\pi T_{\mathrm{d}}}\right), \\
D=\frac{\nu_{\mathrm{F}} v_{\mathrm{F}}^2}{64\pi^2T_\mathrm{d}^2}
\left|\Psi^{\prime\prime}\left(\frac{1}{2}+\frac{\gamma}{2\pi T_\mathrm{d}}\right)\right|.\quad \quad
\end{split}
\end{equation}

\begin{figure}[t]
\label{Fig1}
\begin{center}
\includegraphics[width=8.8 cm]{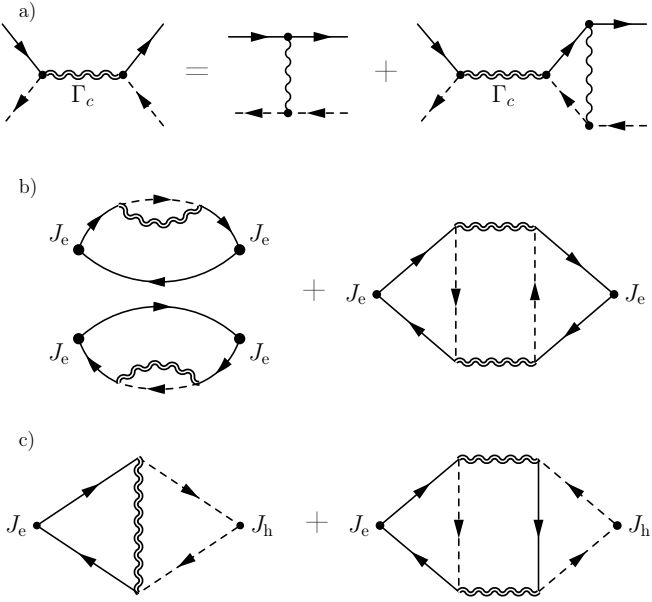}
\caption{(Color online) a) Diagrammatic representation of the Bethe-Salpeter equation for the Cooper propogator $\Gamma_\mathrm{c}$; b) DOS (left) and Aslamazov-Larkin (right) contributions to the conductivity of electron layer. c) Maki-Thompson (left) and Aslamazov-Larkin (right) contributions to transconductivity. Solid (dashed) line corresponds to electrons on the surface of TI film with excess of electrons (holes).}
\end{center}
\end{figure}
\section{Microscopic calculation of conductivity tensor}
Electric currents $\vec{J}_\mathrm{e}$ and $\vec{J}_\mathrm{h}$ in electron and hole surfaces of TI film are connected with the corresponding electric fields $\vec{E}_\mathrm{e}$ and $\vec{E}_\mathrm{h}$ as
\begin{equation}
\label{ConductivityTensor}
\left(
\begin{array}{cc}
J_{\mathrm{e}}\\
J_{\mathrm{h}}\\
\end{array}\right)=\left(
\begin{array}{cc}
\sigma_{\mathrm{e}}& \sigma_{\mathrm{D}}\\
\sigma_{\mathrm{D}}& \sigma_{\mathrm{h}}\\
\end{array}\right) \left(\begin{array}{cc}
E_{\mathrm{e}}\\
E_{\mathrm{h}}\\
\end{array}\right).
\end{equation}
Here $\sigma_{\mathrm{e}(\mathrm{h})}$ are conductivities of the layers and $\sigma_{\mathrm{D}}$ is  transconductivity. For calculation of the contribution of Cooper pair fluctuations to conductivity tensor we use Kubo linear response theory. In that approach the conductivity tensor can be presented in the form
\begin{equation}
\sigma_{\alpha\beta}^\mathrm{R}=\lim_{\omega\rightarrow 0}\frac{\Im[\chi^\mathrm{R}_{\alpha\beta}(\omega)]}{\omega},
\end{equation}
where $\chi_{\alpha\beta}^{\mathrm{R}}(\omega)$ is current-current response tensor that can be obtained by analytical continuation from Matsubara response tensor $\chi_{\alpha\beta}^{\mathrm{M}}(i \Omega_n)$ that is given by
\begin{equation}
\chi_{\alpha\beta}^\mathrm{M}(i\Omega_n)=\frac{1}{2\beta}\int_{-\beta}^\beta e^{i \Omega_n\tau}\langle T_\tau J^x_\alpha(\tau)J_\beta^x(0)\rangle.
\end{equation}
Here $T_{\tau}$ is the time-ordering symbol for a imaginary time $\tau$ and $\Omega_n=2\pi n T$ is a bosonic Matsubara frequency; $J^x_{\mathrm{e}(\mathrm{h})}$ is the electric current operator in the layer with excess of electrons (holes).

Nonzero contribution to the transconductivity comes from the second order diagrams in interlayer Coulomb interaction that is reasonable approximation for weakly coupled bilayer\cite{KamenevOreg}. In the vicinity of the critical temperature Cooper propagator becomes singular and the diagrams with the propagators need to be taken into account. The most divergent diagrams to conductivity of an electron layer (the diagrams to conductivity of a hole layer can be obtained by exchanging the Green functions corresponding to different layers) and transconductivity are presented on Fig.2-b and Fig.2-c. Two DOS diagrams in which Cooper propagator renormalizes single-particle Green function and AL diagram contribute to single layer conductivity. Only MT and AL diagrams contribute to transconductivity and no contribution to it comes form the DOS diagrams.

In the realistic conditions $\gamma\sim T_\mathrm{d}\ll E_\mathrm{F}$ (see discussions in the last Section) the fluctuation contributions to conductivity tensor that are of order of $e^2/h$ is considerable smaller then Drude conductivity $\sigma_{\mathrm{e}(h)}=e^2 E_\mathrm{F}/4h\gamma_{\mathrm{e(h)}}^\mathrm{tr}$. Hence fluctuational contributions to single layer conductivities can be neglected and we conclude that one needs to take into account \emph{only} MT and AL contributions to transconductivity. The primer has been calculated in Ref.[\cite{Hu}] and the later is calculated below. Here we do not omit AL contribution to single layer conductivity since in the next Section we compare results of microscopical and macroscopical approaches.

The AL contribution to the current-current response tensor can be presented in the form
\begin{equation}
\label{ConductivityTensorDiagonal}
\begin{split}
\chi_{\alpha\alpha}^\mathrm{M}(i\Omega_n)=T\sum_{p_n,\vec{q}} \Gamma_{\mathrm{c}}(i\Omega_n+ip_n,\vec{q})\Gamma_{\mathrm{c}}(ip_n,\vec{q}) \\ B_{\alpha}(\vec{q},ip_n+i\Omega_n,i p_n) B_{\alpha}(\vec{q},ip_n, ip_n+i\Omega_n).
\end{split}
\end{equation}
\begin{equation}
\label{ConductivityTensorNonDiagonal}
\begin{split}
\chi_{\alpha\bar{\alpha}}^\mathrm{M}(i\Omega_n)=T\sum_{ip_n,\vec{q}} \Gamma_{\mathrm{c}}(i\Omega_n+ip_n,\vec{q})\Gamma_{\mathrm{c}}(p_n,\vec{q}) \\ B_{\alpha}(\vec{q},ip_n+i\Omega_n,i p_n) B_{\bar{\alpha}}(-\vec{q},-ip_n, -ip_n-i\Omega_n).
\end{split}
\end{equation}
Here we have introduced triangle vertexes
\begin{equation}
\label{TriangleVertex1}
\begin{split}
\vec{B}_{\mathrm{e}(\mathrm{h})}(\vec{q},p_n^{1},p_n^{2})=\pm e v_{\mathrm{F}}\frac{\gamma_{\mathrm{e}(\mathrm{h})}}{\gamma_{\mathrm{e}(\mathrm{h})}^\mathrm{tr}}  T\sum_{w_n,\vec{p}}  G_{\mathrm{h}(\mathrm{e})}(i w_n,\vec{p}) \\ G_{\mathrm{e}(\mathrm{h})}(i w_n+ip_n^{1},\vec{p}+\vec{q}) G_{\mathrm{e}(\mathrm{h})}(i w_n+ip_n^{2},\vec{p}+\vec{q}).
\end{split}
\end{equation}
Here $\gamma_{\mathrm{e}(\mathrm{h})}^\mathrm{tr}$ is the transport scattering rate that comes from the renormalization of current vertexes due to disorder. The main contribution to the sums (\ref{ConductivityTensorDiagonal}) and (\ref{ConductivityTensorNonDiagonal}) comes from the Cooper propagator hence we can neglect dependence of triangle vertexes $\vec{B}_{\mathrm{e}(\mathrm{h})}(\vec{q},p_n^1,p_n^2)$ on frequencies $p_n^1$ and $p_n^2$. After direct calculation we receive
\begin{equation}
\label{TriangleVertex2}
\vec{B}_{\mathrm{e}(\mathrm{h})}(\vec{q},0,0)=2 e D \vec{q} \cdot \gamma_{\mathrm{e}(\mathrm{h})}/\gamma_{\mathrm{e}(\mathrm{h})}^\mathrm{tr} .
\end{equation}
The elements of conductivity tensor can be presented in the following way
\begin{equation}
\label{ConductivityExpression1}
\sigma_{\alpha\beta}^\mathrm{AL}=\frac{\bar{\delta}_{\alpha\beta}\Lambda_{\alpha\beta}}{2}\sum_{\vec{q}}(2eD\vec{q})^2
Y_\vec{q}.
\end{equation}
where the factor $\Lambda_{\alpha\beta}=\gamma_\alpha\gamma_\beta/\gamma_\alpha^\mathrm{tr}\gamma_\beta^\mathrm{tr}$ takes into account difference between scattering rate $\gamma_{\mathrm{e}(\mathrm{h})}$ and transport scattering rate $\gamma_{\mathrm{e}(\mathrm{h})}^\mathrm{tr}$;  $\bar{\delta}_{\alpha\beta}=1$ for $\alpha=\beta$ and $\bar{\delta}_{\alpha\beta}=-1$ in the opposite case. The function $Y_\vec{q}$ is given by
\begin{equation}\label{YFun2}
Y_\vec{q}= \frac{1}{\omega} [\sum_{ip_n} \Gamma_{\mathrm{c}}(i\Omega_n+ip_n,\vec{q})\Gamma_{\mathrm{c}}(p_n,\vec{q})]|_{\begin{smallmatrix}i\Omega_n\rightarrow \omega+i\delta \\ \omega \rightarrow 0\end{smallmatrix}}.
\end{equation}
Performing summation on Matsubara frequencies and analytical continuation we obtain
\begin{equation}
\label{YFunc2}
Y(q)=\int_{-\infty}^{\infty} \frac{d\epsilon}{4\pi T}\left(\frac{\Im\Gamma_{\mathrm{c}}^{\mathrm{R}}(\epsilon,\vec{q})}{\sinh(\epsilon/2T)} \right)^2=\frac{T\gamma_\mathrm{c}}{2\epsilon_{\vec{q}}^3}.
\end{equation}
After substitution of (\ref{YFunc2}) to (\ref{ConductivityExpression1}) and integration we receive \begin{equation}
\label{ConductivityExpression2}
\sigma_{\alpha\beta}^\mathrm{AL}= \bar{\delta}_{\alpha\beta}\Lambda_{\alpha\beta} \frac{e^2}{h} \frac{T \gamma_{\mathrm{c}}}{4a}
\end{equation}
Using explicit expressions for the coefficients in Cooper propagator (\ref{Coefficients}) we receive the final expression \begin{equation}
\label{ConductivityExpression3}
\sigma_{\alpha\beta}^\mathrm{AL}= \bar{\delta}_{\alpha\beta}\Lambda_{\alpha\beta}  \frac{e^2}{16\pi h}\Psi^\prime\left(\frac{1}{2}+\frac{\gamma}{2\pi T}\right) \frac{1}{\ln (T/T_\mathrm{d})}.
\end{equation}
\section{TDGL based approach}
The AL contribution to the conductivity tensor can be evaluated within the macroscopic approach based on time-dependent Ginzburg-Landau (TDGL) equation\cite{AbrahamsTsuneto}. It is well applicable for  description of the dynamics of Cooper pair fluctuations above the critical temperature. We have microscopically derived the TDGL equation for order parameter of electron-hole condensate $\Delta(\vec{r},t)$. If we add Langevin noise $\eta(\vec{r},t)$ to the equation it has the following form
\begin{equation}
\label{TDGL}
\begin{split}
-\gamma_{\mathrm{c}} \left(\frac{\partial\Delta}{\partial t}+i e (\phi_\mathrm{e}(\vec{r},t) -\phi_\mathrm{h}(\vec{r},t))\Delta\right)= \\ = - D \Delta_\vec{r} \Delta + a \Delta +\eta(\vec{r},t).
\end{split}
\end{equation}
Here coefficients $\gamma_{\mathrm{c}}$, $a$ and $D$ coincide with those in the expression for Cooper propagator (\ref{Coefficients});  $\phi_\mathrm{e}(\vec{r},t)$ and $\phi_\mathrm{h}(\vec{r},t)$ are potentials in electron and hole layers. Langevin noise has the correlation function of the white noise $\langle\eta^*(\vec{r},t) \eta(\vec{r}^\prime,t^\prime)=2\gamma_\mathrm{c}T \delta(\vec{r}-\vec{r}^\prime)\delta(t-t^\prime)$. TDGL equation (\ref{TDGL}) can be reduced to the form of Boltzmann equation with help of the Wigner transformation for the condensate order parameter
\begin{equation}
F(\vec{R},\vec{p},t)=\int d\vec{r}e^{i\vec{p}\vec{r}}\Delta^*(\vec{R}+\vec{r}/2,t) \Delta(\vec{R}-\vec{r}/2,t).
\end{equation}
Function $F(\vec{R},\vec{p},t)$ can be interpreted as distribution function of Cooper pair fluctuations and satisfies the Boltzmann-type kinetic equation
\begin{equation}
\label{CooperTDGLBoltzmann}
\frac{\partial F}{\partial t} + e(\vec{E}_\mathrm{e}-\vec{E}_\mathrm{h}) \frac{\partial F}{\partial \vec{p}}=-\frac{2}{\gamma_{\mathrm{c}}}\epsilon_{\vec{p}}(F-F_0(\vec{p})).
\end{equation}
Here $\epsilon_{\vec{p}}=a+D \vec{p}^2$ is energy of Cooper pair and $F_0(\vec{p})=T/\epsilon_{\vec{p}}$ is the distribution function in equilibrium induced by Langevin noise. The full force $\vec{f}_\mathrm{c}=e(\vec{E}_\mathrm{e}-\vec{E}_\mathrm{h})$ acting on Cooper pair is the difference of the forces acting on its components since they are oppositely charged. Electric currents in electron and hole layers carried by the Cooper pair fluctuations are given by
\begin{equation}
\label{CooperCurents}
\vec{j}_{\mathrm{e}(\mathrm{h})}(\vec{r})=\pm e \sum_{\vec{p}}\vec{v}_{\mathrm{p}}\, F(\vec{R},\vec{p},t).
\end{equation}
where $\vec{v}_\vec{p}= \partial \epsilon_{\vec{p}}/\partial{\vec{p}}=2D \vec{p}\,$ is the velocity of Cooper pairs. The system of equations (\ref{CooperTDGLBoltzmann}) and (\ref{CooperCurents}) has the important difference from the analogous one for Cooper pair fluctuations in superconductors \cite{MishhonovIntro}. The force acting on a Cooper pair in a superconductor is $\vec{f}_\mathrm{c}=2e \vec{E}$ and the current is the sum of currents carried by its components since they have the same charge and are not spatially separated.  Calculating currents carried by the Cooper pair fluctuations in electron-hole bilayer in presence of electric fields we obtain
\begin{equation}
\vec{j}_{\mathrm{e}(\mathrm{h})}=\pm \frac{e^2}{h} \frac{T\gamma_\mathrm{c}}{4a} (\vec{E}_\mathrm{e}-\vec{E}_\mathrm{h}).
\end{equation}
The calculated contribution of Cooper pair fluctuations to the conductivity tensor differs from the microscopically calculated one (\ref{ConductivityExpression2}) by the factor $\Lambda_{\alpha\beta}$. If the short-range disorder is dominating scattering mechanism then $\Lambda_{\alpha\beta}=4$ due to suppression of backscattering for Dirac electrons and holes (For other scattering mechanisms see \cite{DasSarmaAdamHwangRossi} and references therein). But if the subtle difference between the scattering rate $\gamma_{\mathrm{e}(\mathrm{h})}$ and the transport scattering rate $\gamma_{\mathrm{e}(\mathrm{h})}^\mathrm{tr}$ is neglected then the microscopic and macroscopic approaches give the same analytical result.
\section{Analysis and discussions}
Transresistance of a bilayer that is the ration between current in active layer to voltage drop in passive layer is measured experimentally. The electric current at side surfaces shunts the layers and interferes with the Coulomb drag effect. The problem can be overcame if the side surfaces are gapped by magnetic doping or by proximity effect to insulating ferromagnet. Recently both mechanisms has been demonstrated experimentally \cite{FerromagnetProximity,MagneticDoping1,MagneticDoping2}.  The side shunting surfaces become also unimportant if the active area of the film in which the contacts are situated and drag effect takes place is considerably smaller then the full area of TI film. In both cases the transresistance is connected with components of the conductivity tensor in the usual way \begin{equation}
\label{Transresistivity}
\rho_{\mathrm{D}}=-\frac{\sigma_{\mathrm{D}}}{\sigma_\mathrm{e} \sigma_\mathrm{h}-\sigma_{\mathrm{D}}^2}.
\end{equation}

According to (\ref{Transresistivity}) fluctuational contributions to conductivity tensor that are of order $e^2/h$ became important for Coulomb drag effect if they are comparable with drag conductivity. Ratio between them and bare value of single layer conductivity that can be approximated by Drude formula $\sigma_{\mathrm{e}(h)}=e^2 E_\mathrm{F}/4h\gamma_{\mathrm{e(h)}}^\mathrm{tr}$ depends on ratio between parameters $\gamma$, $T_\mathrm{d}$ and $E_\mathrm{F}$. Electron-hole Cooper pairing is fragile both to long-range and short-range disorder and $T_\mathrm{d}\ll E_\mathrm{F}$ in weak-coupling limit. Hence in dirty limit $T_\mathrm{d}\ll\gamma\sim E_\mathrm{F}$ the pairing is suppressed. Moreover since predicted critical temperature does not exceed degrees of Kelvin the ultraclean limit $\gamma\ll T_\mathrm{d}$ is very difficult to realize experimentally. Hence we conclude that the \emph{only} regime $\gamma\sim T_\mathrm{d}\ll E_\mathrm{F}$ corresponds to the realistic conditions. In that regime fluctuational contributions are considerably smaller then Drude term and contributions of Cooper pair fluctuations to single layer conductivities can be neglected. Hence we conclude that electron-hole Cooper pair fluctuations do not influence single layer transport. In the realistic conditions the drag conductivity in the denominator of (\ref{Transresistivity}) can be neglected and MT and AL contributions to transresistivity do not interfere. The AL contribution to the transresistance according to the (\ref{Transresistivity}) is given by
\begin{equation}
\label{DragResistivityAslamazovLarkin}
\rho_{\mathrm{D}}^\mathrm{AL}=\frac{h }{e^2} \frac{\gamma_\mathrm{e} \gamma_{\mathrm{h}}}{\pi E_{\mathrm{F}}^2} \Psi^\prime\left(\frac{1}{2}+\frac{\gamma_\mathrm{e}+\gamma_\mathrm{h}}{4\pi T}\right) \frac{1}{\ln (T/T_\mathrm{d})}.
\end{equation}
The formula (\ref{DragResistivityAslamazovLarkin}) is the main result of the work.

The AL contribution to transresistivity diverges in the vicinity of the critical temperature $T_{\mathrm{d}}$ in the critical manner as $\rho_\mathrm{D}^\mathrm{AL}\sim(T-T_\mathrm{d})^{-1}$. At higher temperatures $\Delta T \gtrsim T_\mathrm{d}$ it is decreasing in logarithmical way in the same manner as MT contribution\cite{Hu}. Hence at high temperatures the contributions are undistinguishable. But in the vicinity of the critical temperature the later has has logarithmic singularity and can be approximated as\cite{Hu}
\begin{equation}
\label{DragResistivityMakiThompson}
\rho_{\mathrm{D}}^{\mathrm{MT}}=\frac{h}{e^2} \frac{32 \pi^2 T_\mathrm{d}^2}{35 \Xi(3) E_{\mathrm{F}}^2}  \ln\left(\frac{\pi T_\mathrm{d}}{8(T-T_\mathrm{d})}\right).
\end{equation}
And $\Xi(3)\approx 1.2$  where $\Xi(x)$ is Zeta-function. The result of competition between AL and MT  contributions in vicinity of the critical temperature depends not only on the singularity strength but also on the prefactors in (\ref{DragResistivityAslamazovLarkin}) and  (\ref{DragResistivityMakiThompson}). In the ultraclean limit $\gamma\ll T_\mathrm{d}$ the dominating one is MT contribution due to the corresponding prefactor. But in the realistic regime  $\gamma\sim T_\mathrm{d}$ the AL contribution plays important role.

\begin{figure}[t]
\label{Fig2}
\begin{center}
\includegraphics[width=8.8 cm]{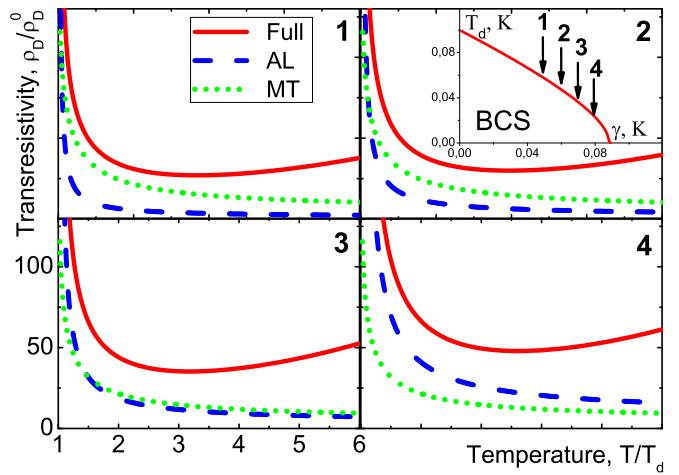}
\caption{(Color online) Temperature dependence of the transresistance (solid red) as well as AL (dashed blue) and MT (dotted green) fluctuational contributions to it. Four subplots have the same scale and correspond to different values of Cooper pair decay: 1) $\gamma=0.05\;\hbox{K}$; 2) $\gamma=0.06\;\hbox{K}$; 3) $\gamma=0.07\;\hbox{K}$; 4) $\gamma=0.08\;\hbox{K}$. The phase diagram of the system, in which BCS denotes the paired state, is depicted on inset of subplot 2. The arrows on it correspond to four subplots.}
\end{center}
\end{figure}

The full value of transresistance includes the term calculated within the second order of perturbation theory in Coulomb interaction $\rho_\mathrm{D}^0(T)$, AL contribution and MT contribution. Drag effect between Dirac layers has been extensively studied in double layer graphene structures \cite{CarregaTudorovskiyPrincipi,HwangSensarmaDasSarma,NarozhnyTitovGornyi,PeresSantosCastroNeto}. In the limit $k_\mathrm{F}\mathrm d\ll1$, which is favorable for electron-hole Cooper pairing\cite{EfimkinLozovikSokolik},   the transresistance of Dirac bilayer is given by \cite{CarregaTudorovskiyPrincipi} $\rho_\mathrm{D}^0= h/e^2\cdot 4\pi T^2/3 E_\mathrm{F}^2 \cdot F(\alpha_\mathrm{c})$, where $F(\alpha_\mathrm{c})$ is the smooth function of the effective fine structure constant $\alpha_\mathrm{c}=e^2/\hbar v_\mathrm{F}$ for Dirac electrons and holes. Here $d$ is topological insulator film width and $k_\mathrm{F}$ is Fermi momentum of electrons and holes. For a calculation of temperature dependence of transresistivity and fluctuational contributions (\ref{DragResistivityAslamazovLarkin}) and (\ref{DragResistivityMakiThompson}) to it we used the following set of parameters $E_\mathrm{F}=5\;\hbox{meV}$ and $\alpha_\mathrm{c}=3.57$, $T_0\approx0.1\;\hbox{K}$, $F\approx 0.05$. The set corresponds to $\hbox{Bi}_2 \hbox{Se}_3$ TI film with width $d=10\; \hbox{nm}$. We also have used four values of Cooper pair scattering rate $\gamma=0.05\;\hbox{K}, 0.06\;\hbox{K}, 0.07\;\hbox{K}, 0.08\;\hbox{K}$ which corresponds to the following critical temperatures  $T_\mathrm{d}\approx0.057\;\hbox{K},0.047\;\hbox{K},0.036\;\hbox{K},0.023\;\hbox{K}$. The ratio $\rho_\mathrm{d}(T)/\rho_\mathrm{d}^0$, where $\rho_\mathrm{d}^0=\rho_\mathrm{d}^0(T_\mathrm{d})$ is transresistance value in second order perturbation theory at critical temperature,  is presented on Fig.2. If $\gamma\lesssim T_\mathrm{d}$ the dominating contribution to the transresistivity is MT one. At  $T_\mathrm{d}\lesssim\gamma \lesssim 4T_\mathrm{d}$ both contributions are important but AL dominates in the vicinity of the critical temperature since it is more singular one. If $\gamma\gtrsim 4 T_\mathrm{d}$ AL contribution dominates the fluctuational transport in the full temperature range. We conclude that AL contribution plays important role within vast region of the phase diagram depicted on inset of Fig.2 and can completely dominate fluctuation transport in electron-hole bilayer.

In a conventional superconductor AL, MT and DOS diagrams contribute to its conductivity in the vicinity of the critical temperature. AL and MT diagrams give positive contribution whereas the DOS ones give the negative one. The result of their competition depends on ratio between the parameters $T_0,E_\mathrm{F},\gamma_\mathrm{el},\gamma_\mathrm{\phi}$. Here $\gamma_\mathrm{el}$ is elastic scattering rate on nonmagentic disorder and $\gamma_\mathrm{\phi}$ is phasebreaking or pairbreaking rate. Since nonmagnetic disorder does not lead to pairbreaking in conventional superconductor (Anderson theorem) there are two parameters connected with it which can considerably differ from each other. In different regimes the result of competition is different and total fluctuational contribution can have as positive, as negative sign. Moreover the number of different regimes can be realized experimentally. So fluctuational transport in superconductors is vast area of condensed matter theory (See \cite{LarkinVarlamov} and references therein). In electron-hole bilayer, on the contrary, the situation is quite definite. There are only AL and MT contributions to transresistivity which have the \emph{same} sign. Any realistic disorder leads to pairbreaking and there is \emph{only} scale $\gamma=(\gamma_\mathrm{e}+\gamma_\mathrm{h})/2$ connected with it. Moreover there is the \emph{only} regime $\gamma\sim T_\mathrm{d}\ll E_\mathrm{F}$ that corresponds to the realistic experimental conditions.

If the fluctuational transport is dominated by AL contribution it can be effectively described within the Boltzmann kinetic equation for fluctuating Cooper pairs derived here. That equation can be easily generalized to the presence of external magnetic field and can used for investigation of the heat transport\cite{UllahDorsey,UssiskhinSondhiHuse}, AC-transport\cite{AslamazovVarlamov} and nonlinear effects\cite{MishonovNonlinear} connected with fluctuating Cooper pairs.

The calculated AL contribution to transconductivity is universal. It depends only on parameters of Cooper propagator and does not depend explicitly on electron and hole single-particle spectrum. Hence we conclude that our theory is well applicable for other realizations of electron-hole bilayer including semiconductor bilayer and double layer graphene structure.

Recently anomalous increasing of transresistance with decreasing of temperature has been measured in semiconductor heterostructure with spatially separated electrons and holes\cite{CroxallExp,MorathExp}. Hence at low temperatures electrons and holes in that system do not behave as weakly-coupled Fermi liquids.  Measured temperature dependence of transresistance contradicts with the predicted\cite{VignaleMacDonald} and experimentalists have tried to interpret the effect as manifestation of the Cooper pair fluctuations. The anomalous dependence has been measured in low-density regime and is not sensitive to concentration mismatch of electrons and holes. So it can be connected with electron-hole pairing not in BCS regime but in the regime of BCS-BEC crossover \cite{Leggett,PieriNeilsonStrinati}. Quantitative theory of the fluctuational drag effect in that regime is interesting and challenging problem. In that system Cooper pairing was predicted to occur in at higher densities of electrons and holes. So if anomalous dependence of transresistanse is found in the regime of high densities in that system then our predictions can be tested.

We have investigated manifestations of fluctuating Cooper pairs formed by electrons and holes populating opposite surfaces of a topological insulator film in Coulomb drag effect.  We have calculated Aslamazov-Larkin fluctuational contribution to transresistivity that is the most singular one and diverges in the critical manner as $(T-T_\mathrm{d})^{-1}$ in the vicinity of the critical temperature. In the realistic conditions $\gamma\sim T_\mathrm{d} \ll E_\mathrm{F}$ it plays important role and can fully dominate the fluctuation transport in electron-hole bilayer. In that case the fluctuational transport can be described within macroscopic approach based on time-dependent Ginzburg-Landau equation developed here. The results can be easily generalized for other realizations of electron-hole bilayer including semiconductor heterostructure and double layer graphene system.

\begin{acknowledgments}
The work was supported by RFBR programs including grant 12-02-31199. D.K.E acknowledge support from Dynasty Foundation. Y.E.L acknowledge support from MIEM and HSE.
\end{acknowledgments}

\bibliographystyle{apsrev}

\begin{thebibliography}{99}
\bibitem{Comment2}
We use here anihilation operators for electrons from both surfaces of topological insulator film for calculations. But for interpretation of results it is more convenient to use the language of electrons and holes.
\bibitem{Comment4}
Introduction of scattering rate to electron and hole Green functions corresonds to the model of uncorrelated potentials acting on them. Taking into account correlation between the disorder's potentials leads only to redifinition\cite{LozovikSokolik} of the paramater $\gamma$ that we treat phenomenologically here.
\end{thebibliography}

\end{document}